\documentclass[useAMS,usenatbib]{mn2e}
\usepackage{graphicx}
\usepackage[update,prepend]{epstopdf}

\title[Multi-component Langmuir modes]{A multi-component Langmuir-mode source for the observed pulsar coherent emission}
\author[P. B. Jones]{P. B. Jones\thanks{E-mail:
p.jones1@physics.ox.ac.uk}  \\
University of Oxford, Department of Physics, Denys Wilkinson Building,\\
Keble Road, Oxford OX1 3RH, U.K.}

\begin{document}

\date{}

\pagerange{\pageref{firstpage}--\pageref{lastpage}}
\pubyear{}

\maketitle

\label{firstpage}

\begin{abstract}
Several classes of neutron star are sources of coherent emission at frequencies of $10^{2}-10^{3}$ MHz: others are radio-quiet.  The primary emission spectra are broadly universal in form over many orders of magnitude in rotation period and polar-cap magnetic flux density.  The existence of nulls and mode-changes in some radio-loud pulsars can be understood only as a manifestation of magnetospheric bistability.  An ion-proton plasma with a possible background of electron-positron pairs is formed at the polar caps of stars with positive corotational charge density and is shown here to be a physical basis for the presence or absence of coherent emission and a likely reason why bistability may be present in the later stages of a pulsar lifetime.

\end{abstract}

\begin{keywords}
instabilities - plasma -stars: neutron - pulsars: general
\end{keywords}

\section{Introduction}

Coherent emission by pulsars usually has a flux density $S(\nu) \propto \nu^{-\alpha}$ in the frequency interval $10^{2} < \nu < 10^{3}$ MHz with a typical spectral index $\alpha \sim 1.8$.  It is remarkable that this is true over intervals of three orders of magnitude in the rotation period $P$ and
five orders of magnitude in the spin-down inferred magnetic dipole field $B_{d}$.  It is also remarkable that the luminosities estimated by Szary et al (2014) from the flux densities $S_{1400}$ for the complete ATNF pulsar catalogue (Manchester et al 2005) are uncorrelated with position in the $P - \dot{P}$ plane.  It is now widely accepted that the source is a plasma instability, possibly plasma turbulence, at low altitudes above the polar caps within the open sector of the magnetosphere.  Exceptional instances have been observed in which the spectral index is $\alpha \sim 0$ or in which the emission is both intermittent and outside the normal well-defined window of pulse longitude.  We shall refer to these as secondary coherent emission and it is possible that new high-sensitivity systems, such as the Square Kilometre Array (SKA) will lead to the discovery of further examples. 

Coherent emission at radio frequencies requires either maser action between suitable energy levels or macroscopic fluctuations in charge and current density.  Unlike incoherent processes such as the emission of X-rays or $\gamma$-rays, the nature of the charged particles contained in these fluctuations is not of immediate significance for the actual radiation process. Emission can depend only on the form of the fluctuations.  The leading incoherent processes must certainly be those of lowest-order quantum electrodynamics involving electrons or positrons. But in the case of coherent emission, there is no requirement that the particles involved should be electrons or positrons. Sources of incoherent emission are, with few exceptions, thought to be at high altitudes in the magnetosphere, near the light cylinder, but are not the subject of this paper.

The existence or otherwise of a suitable plasma instability does, however, depend on the plasma composition and we do not believe that progress can be made without due attention to this.  The relation between rotational angular velocity ${\bf \Omega}$ and polar-cap magnetic flux density ${\bf B}$ is of immediate significance.  For the case of ${\bf \Omega}\cdot{\bf B} > 0$, the polar-cap Goldreich-Julian charge density is negative.  The accelerated particles must be electrons and electron-positron pair creation the only possible secondary process: a reverse flux of positrons has no effect apart from heating the polar cap.

The ${\bf \Omega}\cdot{\bf B} < 0$ case is much more complex. A reverse flux of electrons creates protons which are an important component of the accelerated plasma. Here, we feel obliged to say that empirical methods are not apt for the determination of plasma composition.  The nuclear physics processes are well-established and given their acceptance, there can be no serious doubt that a proton component is present.  As a consequence of the form of the electromagnetic cascade shower produced by reverse electrons, two time constants are introduced, both with observable consequences.

Plasma properties in the ${\bf \Omega}\cdot{\bf B} < 0$ case have been investigated in a number of previous papers (Jones 2010; see also Jones 2013a and papers cited therein). Further papers (Jones 2013b, 2014a,b) based on this work gave reasons for believing that the secondary electron-positron pair plasma which may be formed in pulsars with either sign of ${\bf \Omega}\cdot{\bf B}$ is not the source of the observed primary coherent emission.
Given the ineluctable evidence that the basic spectral properties of the primary coherent radiation are universal, the reluctance to consider sources other than a secondary electron-positron plasma is surprising, although this system might be relevant to processes such as resonant cyclotron emission (Kazbegi, Machabeli \& Melikidze 1991) or cyclotron-Cherenkov resonance (Lyutikov, Blandford \& Machabeli (1999) occurring near the light cylinder.  But neither these sources of light-cylinder coherent emission nor other electron-positron sources such as collective curvature radiation (Istomin, Philippov\& Beskin 2012) and the Langmuir instability in a non-stationary plasma (Asseo \& Melikidze 1998) will be considered further in this paper.

The previous series of papers referred to above makes no specific assumption about the nature of the global magnetosphere and in this respect is at variance with much recent work which uses numerical plasma-physics techniques (for a comprehensive list of published papers see Tchekhovskoy, Spitkovsky \& Li 2013). Model force-free magnetospheres have been constructed extending outwards from a typical inner radius of $0.2 R_{LC}$ with a boundary condition allowing particles of either sign to pass freely through to regions beyond the light cylinder radius $R_{LC}$. In this work, the global magnetosphere (outside the inner radius of the model) determines the current density in the open magnetosphere.  The charge density differs little from the Goldreich-Julian value, but the current density can be significantly different, either larger or of the opposite sign. Such a state can be formed only by the counterflow of oppositely charged particles which, in the ${\bf \Omega}\cdot{\bf B} > 0$ case usually considered, must rely on electron-positron pair creation.  The consequences of a constrained current density for the form of magnetosphere at low altitudes above the polar cap have been investigated recently by Chen \& Beloborodov (2013), Timokhin \& Arons (2013) and Philippov \&Spitkovsky (2014). Pair creation has been modelled in detail by Timokhin \& Arons who adopted the not unusual assumption of an extremely small ($10^{6}$ cm) flux-line radius of curvature.  The present paper is not a critique of this work; it makes the different assumption that the current density is determined at low altitudes above the polar cap. It emphasizes the importance, in the first place, of establishing the physics of the polar-cap region of high energy and particle number densities under conditions such that counterflow at fluxes of the order of the Goldreich-Julian density is not possible, in which the magnetic field ${\bf B}$ is almost precisely immutable, and particle motion is parallel or anti-parallel with it.

The form of global magnetosphere which would exist given the flux of  relativistic particles in the open sector, with moderate Lorentz factors of the orders of $10^{1-2}$, as a boundary condition at $0.2R_{LC}$ is not known.  Observational tests of its nature are not obvious in a normal (middle-aged) pulsar.  Coherent emission is from the polar cap region, levels of incoherent emission from the vicinity of the light cylinder are not usually high, and the spin-down index probably unobservable owing to noise.  In any case, the connection between any of these observables and global magnetospheric structure is by no means direct.

Our proposal here is that the primary coherent emission has a common origin (as its almost universal form appears to indicate) in multi-component Langmuir modes in the magnetospheres of ${\bf \Omega}\cdot{\bf B} < 0$ neutron stars.  This was discussed in Jones (2014b) but that paper had the serious deficiency of failing to describe adequately how the transitions occur in what can be a bistable magnetosphere with nulls and mode changes.
It has since been found (Jones 2014c) that the dielectric tensor for an ion-proton plasma which also includes a background component of electrons and positrons gives a zero Langmuir-mode growth rate for a certain class of electron-positron number densities and spectra.  This can be the case even for densities small compared with the Goldreich-Julian value, and is no more than a consequence of the small electron-proton mass ratio. It appears that this property of multi-component modes allows a modest change in the electron-positron spectrum to extinguish a Langmuir mode which would otherwise have a high growth rate. We shall attempt to describe why such changes can occur naturally, with certain time-scales, in Section 2.2.

A more extensive summary of the plasma properties is contained in Sections 2.1 and 2.2.  The polar-cap system described there is complex with many variables in comparison with the opposite case, ${\bf \Omega}\cdot{\bf B} > 0$, for which Maxwell's equations, the processes of lowest-order quantum electrodynamics, and the laws of mechanics suffice.  Its properties also depend, unfortunately, on two parameters fixed for a given neutron star which are not well known, specifically, the whole-surface temperature $T_{s}$ and the atomic number $Z^{0}_{s}$ of nuclei in the thin surface layer which is the source of ions at the polar cap.  As a consequence of the complexity, the system in its general form has not been completely modelled, but the qualitative results of the limited modelling that has been done are described.

This paper examines a limited question.  Are the properties of the ${\bf \Omega}\cdot{\bf B} < 0$ plasma which are summarized consistent with the
broad properties of the coherent emission and its presence or absence in the various classes of pulsar or isolated neutron star that are observed?  It is not concerned with the \emph{sui generis} cases that are not uncommon in pulsar observations or with more detailed considerations such as the interpretation of polarization which may be intrinsic or the result of propagation through the magnetosphere.  The effects of the latter may be of significance for estimates of the radiation emission height but are not considered here. The question of consistency is addressed in Section 3, with some brief notes on the problems that remain being given in Section 3.6.  Our conclusions are stated in Section 4.  It is hoped that clarification of polar-cap physics will be useful in relation to work on the incoherent processes near the light cylinder which are the source of X-ray and $\gamma$-ray emission.

\section{Physics of the polar cap}

A qualitative understanding of how plasma composition depends on neutron-star parameters is necessary for the purposes described in Section 1.  In order to achieve this, a fairly comprehensive summary of some of the conclusions reached in previous papers (Jones 2013a and earlier papers cited therein) with a statement of the physical factors involved is required and is the subject of this Section.  A clear distinction between negative and positive Goldreich-Julian polar-cap charge densities is necessary. Our summary refers to positive densities unless otherwise stated.

\subsection[]{Plasma composition}

The basis for quantitative calculations has been a circular polar cap of radius $u_{0}$ satisfying space-charge limited flow boundary conditions. Thus there is an atmosphere in local thermodynamic equilibrium (LTE) of which electrons and ions are the dominant components.  Its scale height is of the order of $10^{-1}$ cm, many orders of magnitude smaller than $u_{0}$. A density discontinuity exists between the atmosphere and the solid (or liquid) phase of the surface.  The ion density at the foot of the atmosphere is not well known but is likely to be of the order of $10^{22}-10^{24}$ cm$^{-3}$ .

A reverse flux of electrons from pair creation or photo-ionization must be incident on the polar cap. The electromagnetic showers formed in the atmosphere, or in the atmosphere and solid surface, produce protons at rates which have been calculated and are linear functions of shower energy. The electric field within the LTE atmosphere is determined principally by the electrons and ions and is such that the relatively small density of protons cannot be in static equilibrium: they experience an upward force to the top of the atmosphere.  This fractionation is a direct consequence of the difference in charge-to-mass ratio between ions and protons.  Therefore, the Goldreich-Julian flux is formed preferentially from protons, surplus protons forming a separate atmosphere above the ions. It has an ion component only at times when the proton atmosphere has been exhausted. The diffusion time, from the point of formation in the shower to the top of the LTE atmosphere has been estimated to be of the order of $\tau_{p} \sim 1$ s (see Jones 2013a).

During the lifetime of a pulsar, an extremely thin layer of mass no more than about $10^{-8}$ {\rm M}$_{\odot}$ over the whole surface of the star is sufficient to supply an ion flux at the Goldreich-Julian density.  This is so small that its atomic number $Z^{0}_{s}$ is likely to depend on unknown details of the formation process such as fallback and so must be regarded as an unknown parameter.  The lateral diffusion coefficient for protons, whether derived from the Coulomb collision frequency or from the Bohm mechanism (see Montgomery, Liu \& Vahala 1972) is small, no more than of the order
of $10^{-4}-10^{-2}$ cm$^{2}$ s$^{-1}$, so that the outward proton flux at any point ${\bf u}$ on the polar cap is determined only by the inward electron energy flux at previous times at that point.

Proton (and neutron) production in an electromagnetic shower is almost entirely through the formation and decay of the nuclear giant-dipole state by photons in the $15 - 30$ MeV energy interval.  Consequently, it is localized at the shower maximum which is typically at about $10$ radiation lengths depth.  The properties of a shower at $B \sim 10^{12}$ G are not essentially different from the zero-field case, although at fields greater than the critical $B_{c}= 4.41 \times 10^{13}$ G there are significant differences, among them being the increased importance of the Landau-Pomeranchuk-Migdal effect.  Proton production rates have been estimated in this case but the differences are not large enough to change any of this paper's conclusions..

Photo-disintegration through the giant-dipole state reduces the nuclear charge from $Z^{0}_{s}$ to $Z_{s}$ as ions move upwards to the polar-cap surface. There is a delay between proton formation and its effect on the nuclear charge of ions accelerated from the surface.  This introduces a second time constant of interest, specifically, the time $\tau_{rl}$ in which ions equivalent to one radiation length leave the surface. Both $\tau_{p}$ and $\tau_{rl}$ are consequences of shower structure and both lead naturally to instabilities in the composition of the accelerated plasma.

Ions of nuclear charge $Z_{s}$ have an ionic charge $\tilde{Z}$ in the LTE atmosphere which is dependent on the polar-cap temperature $T_{pc}$. There is a small degree of further ionization $\tilde{Z}\rightarrow Z_{h}$ as ions move initially through the inertial acceleration region (Michel 1974) of the polar-cap blackbody radiation at very small altitudes $z\ll u_{0}$ above the polar-cap where ion Lorentz factors are no more than of the order of unity.
Further acceleration is a consequence of the Lense-Thirring effect (Muslimov \& Tsygan 1992).  But the more significant ionization is at higher altitudes and Lorentz factors in the acceleration region $1 < \eta < \eta_{a}$, where $\eta_{a}$ is a suitable upper limit to the acceleration region.  Photons from a large area of the surface at temperature $T_{s}$ can be boosted in this region by Lorentz transformation to energies sufficient for the ionization of more deeply bound electrons. (Here the polar coordinate is expressed as $r = \eta R$ in terms of the neutron-star radius $R$.)  The final ion charge is $Z_{\infty}$.  Tables of $Z_{\infty}$ and of the mean reverse-electron energy per ion $\epsilon$ have been calculated as functions of the acceleration potential difference $V_{\infty}$ for that ion, treated as an independent variable. These have been obtained for various $B$, $Z_{s}$, and for whole surface temperatures $T_{s} = 1-4\times 10^{5}$ K .

In terms of electrostatics in the corotating frame of reference, photo-electrons accelerated inward to the surface have precisely the same effect as pair creation in partially cancelling the acceleration field.  The electron charge density distribution required to cancel the acceleration field exactly can be found directly from the Lense-Thirring effect expression for the Goldreich-Julian charge density $\sigma_{GJ}(\eta)$ (Muslimov \& Tsygan 1992, see also Harding \& Muslimov 2001).  Its density is,
\begin{eqnarray}
\sigma_{e}(\eta) = -\frac{\kappa}{2\eta^{3}}\frac{\eta^{3}}{\eta^{3} - \kappa}\sigma_{GJ}(\eta),
\end{eqnarray}
and the mean electron track length is $R/2$.  Here, $\kappa \approx 0.15$ is the Lense-Thirring factor which is a function of the neutron-star radius and moment of inertia, but is independent of $B$ and $P$.  Polar-cap modelling (Jones 2013a) has shown that the maximum acceleration field which would be present in the absence of photo-ionization or pair creation is largely cancelled by the reverse electron flux.  Thus it is a fair approximation to assume that the electron track-length distribution is a universal constant.
Photo-ionization transition rates are dependent on the ratio of the threshold photon energy in the ion rest frame to the blackbody photon energy at temperature $T_{s}$.  Hence it follows that ion Lorentz factors and reverse-electron energies are  $\propto T^{-1}_{s}$, approximately.  This is true for the mean values of these quantities even though the modelling shows that the electric field and plasma composition fluctuate over the polar-cap.  This result is of significance in understanding why coherent emission ceases late in the lifetime of a pulsar.

If the magnetic flux density is small, as in the case of millisecond pulsars (MSP), the process is effectively unchanged because zero-field photo-electric cross-sections are adequate.  As we have already mentioned, shower properties are also essentially unchanged.  The role of the magnetic flux density in the process is merely to provide an acceleration potential above the polar cap of a rotating neutron star.  We emphasize that, in this respect, the open magnetosphere described here differs completely from one in which electron-positron pair creation is the essential process and, in particular, from those of ${\bf \Omega}\cdot{\bf B} > 0$ neutron stars.

The state of the polar cap is chaotic as a consequence of the local nature of the shower processes.  This has been confirmed by the construction of a self-consistent discrete-cell polar-cap model (Jones 2013a) based on the tabulated data described above which gave evidence of conal emission and, though limited by the assumption of constant $Z_{s}$,  demonstrated fluctuations in plasma properties not inconsistent with the subpulse modulation that is observed almost universally in the coherent radio emission (Weltevrede, Edwards \& Stappers 2006). But even though $Z^{0}_{s}$ is likely to be a constant over the polar-cap area, there is no reason to assume that the atomic number of nuclei reaching the surface will be other than some function $Z_{s}({\bf u},t)$ of no simple form in either position or time.  In this respect, our modelling has been incomplete.

\subsection[]{Langmuir modes}

The physics considerations summarized above are individually well understood and we regard them as establishing the nature of the accelerated plasma in ${\bf \Omega}\cdot{\bf B} < 0$ neutron stars beyond any reasonable doubt. The generation of coherent emission is extremely efficient (see Jones 2014b) and  the well-defined velocity distribution of an ion-proton system indicates that Langmuir modes should be examined as a possible source.  The longitudinal mode does not couple with the radiation field, but Asseo, Pelletier \& Sol (1990) demonstrated that quasi-longitudinal modes can have similar growth rates and do couple effectively with modes that are capable of leaving the magnetosphere. However, there has since been interest in the generation of plasma turbulence from the longitudinal mode (see Weatherall 1997) and it is not unreasonable to suppose that this is possible through Langmuir-mode growth in a plasma which is not homogeneous and is laterally of finite dimensions.  Under these conditions, we shall assume that coupling with the radiation field does occur,  converting particle longitudinal kinetic energy to radio-frequency emission.

Langmuir-mode angular frequencies for the ion-proton plasma tend to be no more than of the order of the low-frequency turnover that is observed in some pulsar spectra. This is not a problem if we assume (as in Jones 2013b) that the development of turbulence transfers energy to higher wavenumbers as is the case, for example, in homogeneous isotropic fluid turbulence.  However, no quantitative studies appear to have been made for the present case, although the numerical plasma methods of Weatherall (1997, 1998) have shown that, in electron-positron plasmas, very high wavenumbers can be generated such as are observed in the Crab pulsar (Hankins \& Eilek 2007).

The maximum particle longitudinal kinetic energy that can be transferred by any mechanism to electromagnetic radiation is an important quantity.  Luminosities estimated for a group of normal (middle-aged) pulsars by Jones (2014b) showed that the radio-frequency energies generated are typically in the interval $W = 1 - 10$ Gev per unit positive charge in the primary beam, assuming a Goldreich-Julian density. For the ion-proton beam considered here, the kinematic-predicted value of $W$ is given by equation (14) of that paper,
\begin{eqnarray}
W = \frac{xy(\gamma_{p} - \gamma_{I})^{2}}{(1 + y)(x\gamma_{p}+y\gamma_{I})}
m_{p}c^{2},     \nonumber
\end{eqnarray}
where $x = A/\tilde{Z}$ is the mass-to-charge ratio of the ion, $y$ is the mean number of free protons in the beam per unit positive charge of the ions, $\gamma_{I,p}$ are the Lorentz factors and $m_{p}$ is the proton mass.  Its dependence on baryon masses indicates that $W$ can be of the observed order.

Many of the views expressed in this present paper are based on recent studies of Langmuir-mode growth rates for ion-proton plasmas which have an additional small electron-positron component (Jones 2014c). There are several possible sources of pairs other than the curvature radiation (CR) assumed in the classic model of Ruderman \& Sutherland (1975) and in many later papers.  The most quoted mechanism is inverse Compton scattering (ICS) in which a blackbody photon is scattered by an outward moving electron, with magnetic conversion of the resultant photon of high momentum ${\bf k}$ when its transverse component exceeds the threshold $k_{\perp} > 2mc$.  (Directions are with respect to the local ${\bf B}$.) The polarization-averaged photon attenuation coefficient in the classical limit defined by $k_{\perp} \gg 2mc$ and $\chi = k_{\perp}B/2mcB_{c} \ll 1$ is,
\begin{eqnarray}
\lambda_{a} = \frac{m{\rm e}^{2}}{2\hbar^{2}}\frac{B}{B_{c}}\frac{k_{\perp}}{k}
\left(0.377\exp(-4/3\chi)\right),
\end{eqnarray}
(Erber 1966). Conversion lengths $\sim u_{0}$ are given by $\chi \sim 0.1$, which is in the classical region and can be attained for $B \ll B_{c}$. 
The ICS process can be significant for ${\bf \Omega}\cdot{\bf B} > 0$ neutron stars (see Hibschman \& Arons 2001, Harding \& Muslimov 2002) but for the opposite sign, it must be considerably less effective because there is no primary source of outward-moving positrons, with the consequence that any high-energy Compton-scattered photons enter the polar-cap surface without gaining the pair-threshold momentum perpendicular to ${\bf B}$.  Electromagnetic showers also produce neutrons, a small fraction of which may diffuse back to the top of the LTE atmosphere. Therefore, photons from ($n,\gamma$) reactions near the top of the atmosphere are a further source but its importance is difficult to estimate quantitatively. In common with ICS, this process requires a polar-cap field strong enough for single-photon magnetic conversion.  The only source not essentially dependent on $B$ appears to be pair production by photon-photon scattering, but this can be significant only for very high polar-cap temperatures.

The most important result of the study was that the presence of an electron-positron component sparse compared with the Goldreich-Julian number density can reduce the longitudinal Langmuir-mode growth rate to zero if its velocity spectrum overlaps the ion and proton velocities.  Consequently, a small change in the characteristics of a sparse electron-positron component can effectively switch off the turbulence.  An interesting question which then arises is whether or not the whole polar cap is likely to be affected.  It has been shown (see Fig. 1 of Jones 2014c,) that, for a uniform plasma characterized by a given set of parameters, there is a wavenumber of maximum growth rate. But if we regard the wavenumber as an independent variable,  the mode has growth rates of a similar order over an interval that is finite but not unlimited. It is undoubtedly the case that the polar-cap plasma is of finite lateral dimensions and is far from homogeneous in either composition or Lorentz factor. (We assume a zero-potential boundary condition on the cylinder of radius $u_{0}(\eta)$ separating open from closed magnetosphere.)

Under these conditions, what is the state of the polar cap at some general instant of time?  The general free-vibrational state of an inhomogeneous elastic  body would simply be a superposition of its normal modes, though obtaining these does present problems (see, for example, Visscher et al 1991). This suggests that, in the present case, we should consider the growth of a Langmuir mode from an initial fluctuation that is confined to a localized area of the polar cap.  On the basis that growth rates exist over a finite interval of wavenumber in a homogeneous plasma, it is likely that a localized mode would exist with significant amplitude gain on flux lines passing through that area even though it encompasses some variation in plasma parameters.  The general state of the polar cap would then be a superposition of such modes.
If this is accepted, we can see that a modest change in the electron-positron component in some area of the polar cap can effectively switch off a mode.  In this, we can see the basis for both nulls and mode-changes which will be described in the following Section.

\section{A basis for understanding the observed coherent emission}

The polar-cap physics of ${\bf \Omega}\cdot{\bf B} < 0$ neutron stars is necessarily dependent on parameters that are not well known.  These are, principally, the polar-cap surface atomic number $Z^{0}_{s}$ and magnetic flux density $B$, and the whole-surface blackbody temperature $T_{s}$.  It has been assumed here and for modelling purposes that $B$ can be approximated by the ATNF pulsar catalogue values $B_{d}$, there being no alternative (Manchester et al 2005: in modelling the polar cap we have assumed a radius $R = 1.2\times 10^{6}$ cm, which is larger than the ATNF catalogue assumption).  The whole-surface temperature we refer to here is, more precisely, that for a circular area centred on the polar-cap, subtending of the order of a steradian, but with photons in a frame intermediate between the observer and the proper frame of the neutron-star surface.  The most recent survey of existing neutron-star surface temperatures has been published by Keane et al (2013) who also add three upper limits.  Temperatures listed for ages up to $10$ Myr are not inconsistent with the interval of $T_{s}$ used in modelling the polar-cap (Jones 2013a).  There appears to be no direct evidence on the third parameter, $Z^{0}_{s}$, which is not unconnected with $T_{s}$.  Ions with very low $Z_{s}$ have relatively weakly-bound electrons and hence charges $\tilde{Z} = Z_{\infty} = Z_{s}$.  In this case, there would be no photo-ionization and therefore no reduction in the acceleration field and in ion Lorentz factors, which in most neutron stars would reduce the mode growth-rate to insignificant values.  The mode amplitude is $\propto \exp\Lambda(\eta)$, in which,
\begin{eqnarray}
\Lambda = \frac{R\Gamma}{c}\int^{\eta}_{1}\omega^{*}_{z}d\eta^{\prime}
\end{eqnarray}
approximately, at altitude $\eta$, where $\Gamma$ is the dimensionless growth-rate given in Fig. 1 of Jones (2014c) whose value is a slowly-varying function of the relative ion and proton fluxes, and $\omega^{*}_{z}$ is  defined by,
\begin{eqnarray}
\omega^{*2}_{z} = \frac{4\pi N_{z}(\eta)Z_{\infty}^{2}{\rm e}^{2}}{M_{z}\gamma^{3}},
\end{eqnarray}
in terms of the mass $M_{z}$, number density $N_{z}$ and Lorentz factor $\gamma(\eta)$ of the accelerated ion.  In general, the most significant contributions to $\Lambda$ are from small $\eta$, where $\gamma$ is also small.  Assuming a constant $\gamma$, the growth-rate dependence on the basic neutron-star parameters is given by $\Lambda \propto (-B\cos\psi)^{1/2}P^{-1/2}\gamma^{-3/2}$ excluding the parameter $\Gamma$ and the charge-to-mass ratio of the ion in the LTE atmosphere. (Here $\psi$ is the angle between ${\bf B}$ and ${\bf \Omega}$). It has been assumed in previous papers that $\Lambda > 30$ is adequate for the development of turbulence. Thus even if a proton component were present, acceleration of very small-$Z_{s}$ ions to high Lorentz factors would not allow adequate growth of the Langmuir mode.  The processes described in Section 2.2 would not function. But provided this limit is not approached, the value of $Z^{0}_{s}$ is not critical.

\subsection[]{Factors essential for coherent emission}

With these qualifications, we can see qualitatively how the presence or absence of coherent emission depends on $B$ and $T_{s}$.  In comparisons with observation, we can assume only that $T_{s}$ decreases as a function of characteristic age $t_{c} = P/2\dot{P}$. Then the polar-cap modelling in Jones (2013a) and the Langmuir-mode studies (Jones 2014c) reveal two distinct factors.

(i)	   Photo-ionization transitions require black-body photon momenta $k_{bb}$ to exceed the threshold, immediately above which cross-sections are large.
Very approximately, they should satisfy $\gamma k_{bb}c = E_{B}$, where $E_{B}$ is the separation energy of the least bound electron at ion charge
$\tilde{Z}$.  Thus the Lorentz factor required for photo-ionization is
$\gamma \propto T_{pc}T^{-1}_{s}$.	 Hence decreasing $T_{s}$ at constant $B$ requires that ions be accelerated to higher Lorentz factors for photo-ionization, as shown in Section 2.1.  As the limit $T_{s} \rightarrow 0$ is approached, the reverse-electron energy per ion ($\epsilon$) first increases, $\propto T^{-1}_{s}$, as does the rate of proton production.  But if ions have to be accelerated to very high Lorentz factors for the required photo-ionization rate, the mode-amplitude growth exponent $\Lambda$ given by equation (3) will be decreased to the extent that growth-rates in the limit fall to values too small to allow the development of turbulence and observable coherent emission.  The pulsar then enters a permanent radio-quiet state.

(ii)	The rate of ICS pair creation depends on both $B$ and $T_{s}$.  There are two separate reasons. Equation (2) shows that single-photon magnetic conversion at $10^{12}$ G requires a transverse photon momentum $k_{\perp} \approx 9mc$.  But at much higher fields $B \sim B_{c}$, pair creation transition rates are high immediately above the threshold $k_{\perp} = 2mc$. Whilst transverse momentum is not conserved at finite $B$, the final ICS photon $k_{\perp}$ must be balanced by a corresponding component in the Fourier transform of the electron Landau function.  Thus the probability that an ICS black-body photon would reach the threshold for pair creation is small at $10^{12}$ G but increases as $B \rightarrow B_{c}$.  Hence the probability of mode suppression increases rapidly as a function of $B$.  Secondly, the kinematics of ICS in the Klein-Nishina region of energy for an inward-moving electron show that the threshold energy for production of a final-state photon with $k_{\perp} > 2mc$ is $\propto T^{-1}_{s}$.  the blackbody photon density is $\propto T^{3}_{s}$ so that the probability of pair creation at a fixed $B$ is a rapidly increasing function of $T_{s}$.  An increased electron-positron density with the appropriate momentum spectrum tends to suppress one (or more) Langmuir modes and hence the turbulence developing on a given set of flux lines according to arguments given in Section 2.2.

Factors (i) and (ii) determine the conditions under which coherent emission is possible.  But in using them  we have to remember that $B$ may differ significantly from $B_{d}$ and that $Z^{0}_{s}$ is unknown.

Some of the relevant observational data concerning different sets of pulsars are shown in Fig.1. These are not intended to be complete but are representative of the class concerned. On the basis of equation (2) and of the characteristics of curvature radiation, the pair-creation threshold can be approximated by a fixed value of the variable $X = B_{d12}P^{-7/4}$, assuming a dipole field (see also Fig. 1 of Harding \& Muslimov 2002). A threshold value $X_{c} = 6.5$ is consistent with Harding \& Muslimov.  

Normal pulsars are too numerous to be shown individually: the vertical broken lines show the half-maxima of their distribution in $B_{d}$ and most lie at $X < X_{c}$.  Pulsars exhibiting nulls and nulls plus mode changes are shown separately, and are those listed in Tables 1 and 2 of Wang, Manchester \& Johnston (2007).  Radio-loud millisecond pulsars (MSP) are also shown in separate groups: those listed in the second Fermi LAT catalogue (Abdo et al 2013) and those which have no observable $\gamma$-emission but are contained in Table 3 of Kramer et al (1999). The Rotating Radio Transients (RRAT) are those listed with timing solutions by Keane et al (2011). The Central Compact Objects (CCO) are those described by Gotthelf et al (2013): the X-ray Isolated Neutron Stars (XINS) and High-Field Rotation-Powered Pulsars (HBRPP) are those listed by Kaplan \& van Kerkwijk (2009). The magnetars are from the McGill catalogue (Olausen \& Kaspi 2014).

\subsection[]{Normal pulsars}

Pulsars not placed in the specific classes listed here are referred to as normal: they are radio-loud and comprise the major part of the ATNF catalogue.  The typical field is $B\sim 10^{12}$ G so that the classical limit for single-photon magnetic conversion is apt. A small fraction of them lie at $X > X_{c}$ and the immediate question is whether in these there is self-sustaining CR electron-positron pair creation at the polar cap.  Any having ${\bf \Omega}\cdot{\bf B} > 0$ should produce pairs but our view (Jones 2013b, 2014b) is that they would not be radio-loud in the sense of having the usual large negative spectral index.  It is possible to argue that a state of pair formation at $X > X_{c}$ in those with ${\bf \Omega}\cdot{\bf B} < 0$ would be unstable against a transition to the ion-proton state.  Ion emission with formation of reverse photo-electrons must always be possible in polar-cap areas near the boundary $u_{0}$ where the acceleration field is not large enough to support CR pair creation. Any inward fluctuation of this zone must further reduce the acceleration field and so enlarge its area.  For this reason, we believe that an ion-proton plasma is the source of coherent emission even in very young pulsars.  Polar-cap modelling (Jones 2013a) has shown, unsurprisingly, that relatively high surface temperatures, $T_{s}\sim 5\times 10^{5}$ K, much reduce both the mean acceleration potential difference above the polar cap and fluctuations away from this state, which is therefore effectively stable.

Consideration of factors (i) and (ii) for normal neutron stars at $X < X_{c}$ with ${\bf \Omega}\cdot{\bf B} < 0$ and $B_{d} \sim 10^{12}$ G shows that ICS pair production is negligible compared with the ${\bf \Omega}\cdot{\bf B} > 0$ case. Also the polar-cap temperature $T_{pc}$ is not high enough to give significant rates for pair production by photon-photon scattering.  Thus for the ${\bf \Omega}\cdot{\bf B} < 0$ case, with moderate values of $B$, a radio-loud state should be expected until the final turn-off.  Observable coherent emission ends because the amplitude growth rate exponent,
$\Lambda \propto B^{1/2}P^{-1/2}T^{3/2}_{s}T^{-3/2}_{pc}$, derived from equations (3) and (4), must eventually become small as $T_{s}$ declines.  Here, we neglect $\cos\psi$, also any $\psi$-dependence in our use of the ATNF catalogue values of $B_{d}$.

The heat flux represented by $T_{pc}$ has two sources: the reverse-electron energy and the heat loss from the interior of the star which is the source of $T_{s}$.  The time-averaged reverse-electron energy flux has an upper limit equal to that which would create a Goldreich-Julian flux of protons (see Jones 2010; equation 33).  The shower energy required to produce one proton is a slowly-varying function of $B$ but is treated here as a constant. For high surface temperatures exceeding $\sim 100$ eV, it is easy to confirm that for a normal star the heat flow from the interior exceeds that of the reverse flux.  Then $T_{pc} = T_{s}$ is a fair approximation.  But the eventual decline of $T_{s}$ and the increase in $P$ will  reduce $\Lambda$ so that growth becomes too small for the development of turbulence.

The presence of $B$ in equation (3) derives from the Goldreich-Julian charge density: it is therefore the true magnetic flux density, not $B_{d}$. In principle, $B = \zeta B_{d}$ with $\zeta$ larger or smaller than unity: an example would be the field of a non-central dipole.  Therefore in Fig. 1,
which gives $B^{1/2}_{d12}P^{-1/2}$ as a function of $B_{d}$ for the various groups considered here, it should not be expected that a constant $B$-independent cut-off value will be seen.

In fact, it is possible to superimpose on Fig.1 a curve showing a well-defined cut-off over the first five orders of magnitude in $B_{d}$.  The multitude of normal pulsars in the ATNF catalogue is too large to be shown, but the relevant boundary of their distribution has a sharp cut-off shown in the diagram as an arc of a curve.  It would be consistent with the imaginary superimposed curve referred to above, were we to draw it.  The expression for $\Lambda$ does not necessarily apply for $B_{d} > B_{c}$ because mode growth-rates are liable to be reduced to zero by pair creation, as described in Section 3.1 (ii).

The indicated cut-off value of $B^{1/2}_{d12}P^{-1/2}$ is not precisely independent of $B_{d}$ but varies remarkably slowly over the five orders of magnitude to which we refer.  The expression for $\Lambda$ has been obtained by the simplest scaling considerations and too much should not be expected of it.  It is also possible that there are systematic rather than random variations in $\zeta$.  Thus $\zeta > 1$ at small $B_{d}$ (the MSP and CCO) would render the true cut-off less $B$-dependent.  It may be relevant that most MSP are in binaries, and their history is quite different from that of the normal multitude.

\begin{figure*}
\includegraphics[trim=10mm 30mm 30mm 70mm,clip,width=168mm]{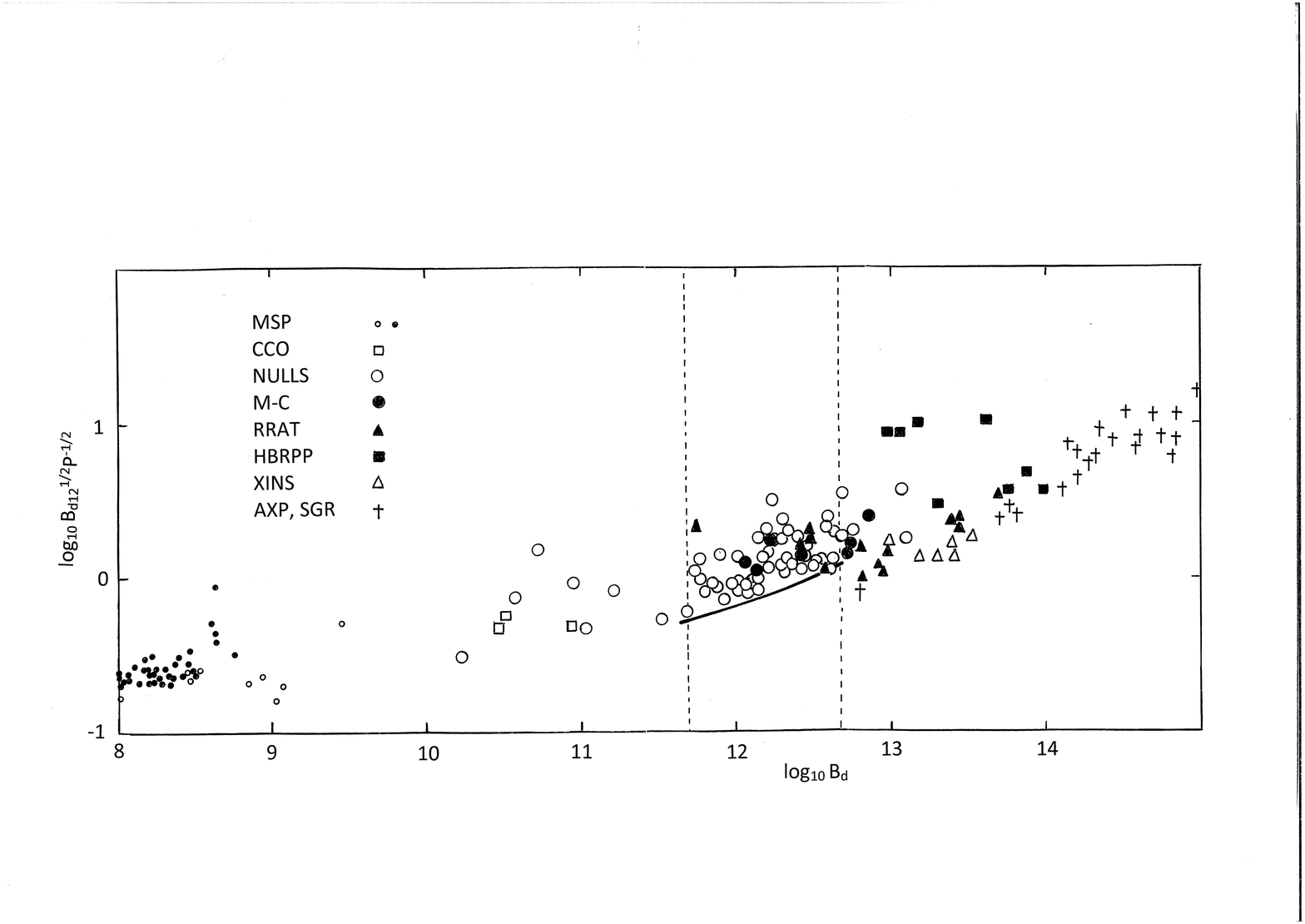}

\caption{The factor $B^{1/2}_{d12}P^{-1/2}$ is shown as a function of $B_{d}$ for the sets of pulsars and neutron stars listed in the final paragraph of Section 3.1.  These are not in all cases complete lists of their class but are convenient representative groups.  The large number of normal pulsars in the ATNF catalogue are not shown individually but vertical broken lines show the half-maxima of their distribution in $B_{d}$.  From the ATNF catalogue, it has been possible to deduce the lower boundary of their distribution and this is shown as a solid arc between the half-maximum limits.  The MSP are divided into those with a Fermi LAT catalogue entry (solid circles), all of which are radio-loud, and a radio-loud set with no observed $\gamma$-emission.  Pulsars showing nulls and nulls plus mode changes are also shown separately.}

\end{figure*}

We have noted previously, in Section 2.1, that the fluctuations in polar-cap acceleration potential and plasma composition that are the result of the time constant $\tau_{p}$ are not inconsistent with the sub-pulse modulation which is observed almost universally.  But it is likely that some of the modulation, particularly that observed at high time-resolution, is a consequence of the turbulent plasma state we believe to be the radiation source.

Apart from the papers of Weatherall, which considered pair plasma, there appears to be no work on radiative emission from the proton-ion turbulence we assume is formed. The nanosecond structure observed by Hankins \& Eilek (2007) in the Crab pulsar main pulse provides possibly the most exacting test of any emission model.  The example shown in Fig. 5 of their paper represents a nanopulse with an integrated flux of $2\times 10^{-20}$ erg cm$^{-2}$. If the source at distance $D$ is moving toward the observer with Lorentz factor $\gamma_{s}$ and, in its rest frame, emits into solid angle $\delta\Omega_{s}$, the direction of emission in the observer frame can be at an angle of the order of $\gamma^{-1}_{s}$ with respect to ${\bf B}$ and the area illuminated is then $\sim D^{2}\delta\Omega_{s}/\gamma^{2}_{s}$. Given any reasonable assumptions about the two unknown parameters here, this is a remarkably large energy.  Let $\lambda_{s}$ be the central wavelength emitted in the rest frame of the source. Collective particle motion and acceleration is parallel with ${\bf B}$.  Thus within the turbulent volume, a basic unit in the source rest-frame giving coherent emission might be a very short-lived quasi-dipole source, axis parallel with ${\bf B}$ and of length $\lambda_{s}$, width $w\lambda_{s}$ and depth $\lambda_{s}/2\pi$. The energy available, per unit positive charge in the ion-proton plasma, for conversion to radio-frequencies, has been estimated previously (Jones 2014b) to be $W \sim 7$ Gev.  The energy that can be extracted from this unit in principle is therefore no more than
\begin{eqnarray}
\frac{w\lambda^{3}_{s}}{2\pi\gamma_{s}}\frac{WB(\eta)}{P{\rm e}c},
\end{eqnarray}
in the observer frame.  The coherent source volume can be large for $w \gg 1$, but the emission solid angle is then reduced to $\delta \Omega_{s}\sim 2\pi/w$.  The insertion of parameter values for the Crab, with $\gamma_{s} = 10$, $\eta = 10$, and the central frequency of $9.25$ GHz, gives $\lambda_{s} = 32$ cm and $w = 9\times 10^{5}$.  The total width $w\lambda_{s} = 3\times 10^{7}$ cm is to be compared with an open magnetosphere radius $u_{0}(\eta) = u_{0}(10) = 2.8\times 10^{6}$ cm.  Clearly, some organization of coherence more subtle than our elementary example would be required.  It must also satisfy the condition that to produce a pulse length of $\delta t = 0.4$ nanoseconds, the units must lie within a height interval of less than $2\gamma^{2}_{s}c\delta t$ in the observer frame.  Thus formation of nanosecond structures is possible in principle, although we do not attempt here to estimate their probability. But it must be obvious how much more difficult it would be to understand these structures in terms of a light cylinder source.

\subsection[]{Millisecond pulsars}

Two sets of MSP are shown in Fig. 1; those listed in the Fermi LAT catalogue (Abdo et al 2013) and those not listed in that catalogue but present in Table 3 of Kramer et al (1999). A notable feature is the absence of radio-quiet MSP (defined as a flux density $S_{1400} < 30$ $\mu$ Jy) in the Fermi LAT catalogue. This is complicated by the different detection procedures used; either known radio ephemerides or blind searches using periodicities in LAT data. Nevertheless, Romani (2012) has estimated that the fraction of radio-quiet MSP can be no more than $\sim 1/3$.  In view of the very small rotation periods and magnetic fields it is notable that the MSP radio spectra in both sets are very similar to those of normal pulsars (Jenet et al 1998, Kramer et al 1999, Espinoza et al 2013).

A further unresolved question concerns a small group of 4 MSP in which the radio and $\gamma$-emission light curves are in phase (Venter, Johnson \& Harding 2012).  The Crab pulsar is the only other known example of this phase-alignment: phase leads or lags are otherwise the rule.   Following the outer-gap model introduced by Cheng, Ho \& Ruderman (1986) and further developed by Cheng, Ruderman \& Zhang (2000) to include pair creation by photon-photon collisions, it is now generally assumed that the source of $\gamma$-emission must be close to the light cylinder.  Phase-alignment has been used as evidence for coincident source regions, but this is open to question.

  With regard to the Crab pulsar, we have commented in Section 3.2 that it is not obvious how the nanosecond structure in its main pulse (Hankins \& Eilek 2007) can be reconciled with emission from the vicinity of the light cylinder.  These indicate a polar-cap source for the coherent radio emission.  If this is so, and the Crab is an ${\bf \Omega}\cdot{\bf B} < 0$ neutron star, the source of its generally accepted pair production must be an outer gap.  Pairs could certainly be formed at the polar cap of an ${\bf \Omega}\cdot{\bf B} > 0$ Crab pulsar but it would be radio-quiet or have emission very different from that observed.
  
The growth-rate exponent described in Section 3.2 can also be evaluated for MSP under the assumption that $T_{pc} = T_{s}$.  The distributions of $\Lambda$ for the two sets of MSP are each quite compact: for the Fermi LAT set the mean value is, $\langle B^{1/2}_{d12}P^{-1/2}\rangle = 0.28$ and for the Kramer et al set, $0.24$.  These low values in radio-loud MSP might be thought a problem, but as noted in Section 3.1, the magnetic flux density in the expression $\Lambda \propto B^{1/2}P^{-1/2}$ is the real (unknown) value, not $B_{d}$.  Also, MSP almost certainly have a history different from the typical isolated pulsar, making the implicit assumption of a common $Z^{0}_{s}$ doubtful.
  
  Kramer et al find that their MSP observations are consistent with a single emission altitude and a compact source, equivalent to a light travel distance of no more than $2.4\times 10^{5}$ cm.  Pairs are not made at the polar cap of an MSP for the reasons given in Section 2.2 and in (i) of the Section 3.1.  The plasma in the ${\bf \Omega}\cdot{\bf B} < 0$ case has a negligible electron-positron component; there is no mode suppression and therefore nulls should not be a feature of MSP.

But there is no reason to suppose that MSP with
${\bf \Omega}\cdot{\bf B} > 0$ should not exist. (The Fermi LAT catalogue contains a number of non-MSP sources that are radio-quiet.)  The problem is why are radio-quiet MSP not seen?  In view of the estimate made by Romani (2012), this remains unresolved.

The MSP magnetospheric structure near the light cylinder is not well-established but we can comment that electrons must pass through the light cylinder in the ${\bf \Omega}\cdot{\bf B} < 0$ case with an integrated flux of the order of the Goldreich-Julian magnitude.  The reason is that there exists the return-current problem: the star and magnetosphere inside the light cylinder must maintain neutrality or a constant net charge over time intervals of the order of the rotation period.  Although this paper is not directly concerned with $\gamma$-emission, an outer gap sustained by photon-photon collisions should be possible for either sign of ${\bf \Omega}\cdot{\bf B}$.

\subsection[]{Nulls mode-changes and RRAT}

This class of activity is clearly a feature of increasing polar magnetic flux density, as demonstrated in Fig.1.  In the case of nulls, observable radio emission ceases abruptly, within a time interval shorter than the rotation period, although in some cases very weak but detectable emission remains (Esamdin et al 2005, Wang et al 2007): mode-changes are similarly prompt. 

Pulsars listed in Tables 1 and 2 of Wang et al consist of 63 that have a finite null fraction, of which 5 also exhibit mode changes.  There are also 2 showing mode changes but with no measured null fraction.  The mean value of $B$ is significantly larger than for the whole population and this is apparent in Fig. 1 in which those with mode changes are shown separately.   The age distribution extends typically over the interval $4 - 40$ Myr and there is some evidence (see Wang et al) that the null faction increases with age.

Mode changes are much less well catalogued, but in a broad sense, nulls and mode-changes appear as different aspects of the same phenomenon.  A specific pulsar which exemplifies this is B0943+10 (Hermsen et al 2013, Mereghetti et al 2013, Bilous et al 2014).  The mechanism underlying this phenomenon follows directly from Sections 2.2 and Section 3.1. We suggest that it is simply the suppression of one (or more)  Langmuir modes by a change in the density or momentum spectrum of the electron-positron component on flux lines emerging from some section of the polar-cap area, whilst other modes remain and are observable. There are two obvious characteristic times which can be significant: the first is $\tau_{p}\sim 1$ s which is the basic time-scale for fluctuations in electric field and plasma composition at the polar cap.  The second is $\tau_{rl}$, the time in which ions equivalent to one radiation length are accelerated from the polar-cap atmosphere,
\begin{eqnarray}
\tau_{rl} \approx 2.1\times 10^{5}\left(\frac{-P\sec\psi}{Z_{s}B_{12}
\ln(12Z^{1/2}_{s}B^{-1/2}_{12})}\right) \hspace{4mm}{\rm s}
\end{eqnarray}
where $\psi$ is the angle between ${\bf \Omega}$ and ${\bf B}$.  This  represents the medium time-scale of fluctuations in $Z_{s}$ (see Jones 2013a).  It is a consequence of the fact that the reduction in nuclear charge from $Z^{0}_{s}$ to $Z_{s}$ by giant-dipole state formation and decay occurs at a limited interval of depth near the shower maximum.  The change in $Z_{s}$ is unlikely to be as rapid as the observed onset of a null, but the change in the exponent $\Lambda$  given by equation (3) need not be large to suppress the amplitude of the mode concerned.

Pair production by photon-photon scattering is not essentially $B$-dependent, but the threshold momentum for an incoming curvature-radiation photon is $k_{c} = 2m^{2}c^{2}/(k_{bb} + k_{bb\parallel})$ and is most unlikely to be reached. There is no evidence that values of either $T_{pc}$ or $T_{s}$ are large in pulsars exhibiting nulls or mode changes.

Nulls of very short duration, usually a single period, are frequent but we believe them to be no more than a direct consequence of the fluctuations with time-constant $\tau_{p}$ that are a feature of the model polar cap (Jones 2013a).  Times given by equation (6) are not inconsistent with what are termed medium time-scale nulls and with mode changes. A null fraction increasing with age is clearly related to the age-related decrease in $\Lambda$.  The cut-off value $\Lambda_{c}$ is not necessarily well-defined because it must depend on other factors which have been neglected in the scaling expression for $\Lambda$.  An example is $Z_{s}$ which can change in a particular region of the polar cap as a consequence of the medium-term instability with time-scale $\tau_{rl}$.

Our final comments in this Section concern the RRAT.  These have an age distribution which differs little from the Wang et al nulls.  But the magnetic fields are an order of magnitude larger.  They are viewed here as merely an extreme case of the nulling phenomenon.  They are radio-quiet except for intervals of one or possibly two pulses, which are of the order of $\tau_{p}$ in length.  We believe it is a natural consequence of the increased field that, for the reasons we have already described previously in Section 3.1, the Langmuir modes are suppressed except for short time intervals of the order of $\tau_{p}$.  They appear to have no observable successor population in the ATNF logarithmic $P-\dot{P}$ diagram and, as we would expect on the basis of Section 3.1, are presumably evolving towards a radio-quiet state.

\subsection[]{HBRPP, CCO, XINS, AXP and SGR}

These groups of neutron stars are presently observed to be radio-quiet except for the HBRPP and the AXP J1810-197.  The latter source does have coherent emission (Serylak et al 2009) but not with the usual spectral properties, and falls into the class we have defined in Section 1 as secondary emission.
The HBRPP are defined as those with $B_{d} > 10^{13}$ G.  All groups except the Central Compact Objects (CCO) are positioned on the right-hand side of Fig. 1.  Surface temperatures are typically $\sim 100$ eV for the XINS, $100 - 200$ eV for HBRPP (Kaplan \& van Kerkwijk 2009).  The AXP and SGR have much higher values, typically $\sim 300$ eV.

An immediate question must be why are HBRPP radio-loud but the XINS, with similar values of $B_{d}$ and $T_{s}$, radio-quiet?  Some caution is necessary here because the XINS are small in number and have been detected by their $4\pi$ X-ray emission: narrow radio-frequency beams may have been missed.  It is also possible that they are all ${\bf \Omega}\cdot{\bf B} > 0$ neutron stars.  But for present purposes we shall discount these possibilities.

Reference back to Section 3.1 indicates that coherent emission has two requirements.  The mode growth-rate given by equation (3) must be sufficiently large to produce non-linearity and turbulence, and mode suppression through pair creation at large $B$ must be absent.  The growth-rate exponent reduces to $\Lambda \propto B^{1/2}P^{-1/2}$ for the approximation $T_{pc} = T_{s}$, following the discussion given in Section 3.2.  Typical HBRPP surface temperature are in the interval $100 - 200$ eV and justify this approximation.  The distributions of $B$ for the two groups are very similar. Then for the XINS and for the HBRPP listed as radio-loud by Kaplan \& van Kerkwijk, the values of $\Lambda$ fall into two well-separated  groups. The average values are $\langle B^{1/2}_{d12}P^{-1/2}\rangle_{HBRPP} = 7.4$ and  $\langle B^{1/2}_{d12}P^{-1/2}\rangle_{XINS} = 1.6$.  This is a large difference in an amplitude exponent for which we have assumed that a value $\Lambda = 30$ gives adequate gain, and appears a plausible explanation for the presence or absence of coherent emission. Our only reservation here is that a common value of $Z^{0}_{s}$ is impicit in this growth-rate comparison.  The two groups may have different histories, with different surface properties as a consequence.

We have also to comment on the CCO, in particular, the group of J0821-4300, J1210-5226, and J1852+0040 observed by Gotthelf et al (2013).  Here, the assumption $T_{pc} = T_{s}$ appears satisfactory, and the values of $B^{1/2}_{d12}P^{-1/2}$ are, 0.51, 0.48, and 0.54 respectively.  These appear to lie just above the notional cut-off curve that we referred to in section 3.2 but, as in the case of the MSP, the expression requires $B$ and not $B_{d}$.  Thus the lack of sufficient mode growth may explain the absence of coherent emission now and in the future, but it remains possible that the CCO are ${\bf \Omega}\cdot{\bf B} > 0$ neutron stars.

The magnetars (AXP and SGR) have much larger values of $B$ and of $T_{s}$.
Pair creation in the special case of $B > B_{c}$ has been described by Thompson (2008).  But in these groups, the value of $\Lambda$ given by equation (3) is immaterial.  Following the arguments given in Section 3.1 (ii), the pair creation rates are much larger than in any other group.  Mode growth is not possible.  The magnetars have no obvious radio-loud successors present in the ATNF logarithmic $P - \dot{P}$ diagram and we have to presume that, with cooling, mode suppression remains until $\Lambda$ is too small to produce coherent emission.

\subsection[]{Neutron stars with ${\bf \Omega}\cdot{\bf B} > 0$ and unresolved problems}

There appears to be no reason why neutron stars with ${\bf \Omega}\cdot{\bf B} > 0$ should not exist.  Many authors have assumed them to be the canonical pulsar.  But it has proved difficult to relate the properties of a secondary electron-positron plasma to the phenomena that are observed.  Specifically, the spectral properties of the emission are essentially unchanged over three orders of magnitude in $P$ and five orders of magnitude in $B_{d}$, whereas pair creation in many pulsars through conversion of curvature radiation appears impossible unless an \emph{ad hoc} flux-line curvature is assumed (see Hibschman \& Arons 2001, Harding \& Muslimov 2002).  In some recent LOFAR observations (Hassall et al 2012) emission heights and source size do not correspond with what would be expected for an electron-positron plasma (Jones 2013b).

The view of this paper is that the primary emission arises from the more complex plasma of ${\bf \Omega}\cdot{\bf B} < 0$ neutron stars.  In these, the polar magnetic flux density has no essential part in the emission process except for the generation of the acceleration field and in suppression of the Langmuir mode  at large $B$.
Neutron stars with ${\bf \Omega}\cdot{\bf B} >0$ may be sources of secondary emission, possibly observable through the SKA, but otherwise emit only in the incoherent regions of the spectrum. Young neutron stars with ${\bf \Omega}\cdot{\bf B} > 0$ are sources of X-ray and $\gamma$-ray emission from the vicinity of the light cylinder but are expected to be radio-quiet and to remain so.  A considerable fraction of Fermi LAT pulsars are in this category.  Of $77$ pulsars not classified as MSP, Abdo et al list $35$ as radio-quiet. The radio-quiet neutron-star groups in this Section could have either sign of ${\bf \Omega}\cdot{\bf B}$.

Two obvious unresolved problems remain within Sections 3.3 and 3.4.  We are unable to comment here on the absence of radio-quiet MSP in the Fermi LAT catalogue, but there is some obligation to mention the well-established properties of long-term nulls in three pulsars; B1931+24 (Kramer et al 2006), J1832+0029 (Lorimer et al 2012) and J1841-0500 (Camilo et al 2012).  Whereas the time constants $\tau_{p}$ and $\tau_{rl}$ have values consistent with possible relationships with short and medium-term nulls, times of the order of $10^{7}-10^{8}$ s appear unrelated to any physical quantity, except possibly times derived from the transverse diffusion coefficient of protons or ions in the polar-cap LTE atmosphere.  A further property of these pulsars is that their spin-down torque is approximately halved during their null intervals.  Although measurements of this kind are not possible in most pulsars, it is known that large changes are not universal.  In one case (B0823+26; Young et al 2012) the fractional change in spin-down torque has an upper limit of $0.06$.

The likely explanation for this behaviour is that the flux and nature of particles crossing the light cylinder change during the null. The transition between ion-proton and electron-positron plasma is an obvious example.  Our argument in Section 3.2 that a state of pair production in an ${\bf \Omega}\cdot{\bf B} < 0$ neutron star at $X > X_{c}$ would always be unstable against transition to an ion-proton plasma might be incorrect.  If this were so, such transitions could be significant because all three pulsars have $X \sim X_{c}$.  It is by no means clear which state would produce the larger spin-down torque.

\section{Conclusions}

The ideas summarized in Section 2.1 are based on well-established nuclear physics and in our view there can be no serious doubt that an ion-proton plasma is produced at the polar cap of an ${\bf \Omega}\cdot{\bf B} < 0$ neutron star. The motivation for the series of papers cited there was the belief that the problem of plasma composition must be solved first.  Section 2.2 summarizes ideas based on this which might be more contentious.  The development of turbulence in the inhomogeneous system described there is a formidable problem.  The only simple feature is the $\delta$-function form of velocity distribution which at once suggests Langmuir modes.  The question of whether prior damping of the Langmuir mode might preclude the development of turbulence has not been addressed in detail, although there appears to be no obvious mechanism.

The necessary involvement of condensed matter leads to complexity and, in particular, to a dependence on parameters that are now, and may remain in the immediate future, not well-known.  Almost nothing is known directly about the surface atomic number $Z^{0}_{s}$ and there is the difficulty that our expression for the growth-rate exponent $\Lambda$ requires the actual $B$ which may differ from $B_{d}$.  Most MSP are in binaries, with a history very different from isolated neutron stars.  Nothing is known of the history of the XINS.  The surface temperatures $T_{s}$ in normal pulsars are usually too low to measure, the black-body photons being in an energy interval strongly absorbed by the interstellar medium.  With the complexity of the model, these factors restrict the nature of the predictions that can be made.

The aim of this and previous papers has been to provide a clear physical basis for understanding the nature of the pulsar inner magnetosphere, but only at the broadest level and in a qualitative way.  Given the nature of the system, it may prove difficult to do more in the foreseeable future.  Nonetheless, within these limitations, the work has provided an understanding of how a plasma is created that has the capacity to generate coherent  emission over the very large intervals of $B$ and $P$ that are observed, and how that emission  eventually ceases. The finding that relatively small (background) fluxes of electrons and positrons can affect mode growth has proved very significant.  It is likely to be the basis for the magnetospheric changes that are seen as nulls and mode changes.  It also demonstrates why some very active neutron stars have no observable coherent emission.
More observational evidence on source sizes and altitudes would be of interest, as would high-sensitivity SKA studies of what we have referred to as secondary coherent emission.  An understanding of the inner magnetosphere region of high energy densities may be an essential first step towards achieving a more complete understanding of the whole magnetosphere.

\section*{Acknowlegments}

The author thanks the anonymous referee for comments which lead to a much improved presentation of the contextual nature of this paper.

\bsp

\label{lastpage}

\end{document}